\documentclass[11pt,a4paper]{article}

\usepackage{amsmath,amssymb,amsfonts,mathrsfs}
\usepackage{epsfig,epstopdf}
\usepackage{graphicx,graphics}
\usepackage{color}
\usepackage{float}
\usepackage{ifpdf}
\usepackage[colorlinks=true]{hyperref}
\usepackage{braket}

\newtheorem{theorem}{Theorem}[section]

\newtheorem{corollary}{Corollary}[section]

\newtheorem{definition}{Definition}[section]
\newtheorem{example}{Example}[section]

\newtheorem{lemma}{Lemma}[section]

\newtheorem{problem}{Problem}[section]
\newtheorem{proposition}{Proposition}[section]
\newtheorem{remark}{Remark}[section]
\numberwithin{equation}{section}

\newcommand{\bthm}{\begin{theorem}}
\newcommand{\ethm}{\end{theorem}}
\newcommand{\blem}{\begin{lemma}}
\newcommand{\elem}{\end{lemma}}
\newcommand{\bex}{\begin{example}}
\newcommand{\eex}{\end{example}}
\newcommand{\bprop}{\begin{proposition}}
\newcommand{\eprop}{\end{proposition}}
\newcommand{\bplm}{\begin{problem}}
\newcommand{\eplm}{\end{problem}}
\newcommand{\bmrk}{\begin{remark}}
\newcommand{\emrk}{\end{remark}}
\newcommand{\bdfn}{\begin{definition}}
\newcommand{\edfn}{\end{definition}}
\newcommand{\bcor}{\begin{corollary}}
\newcommand{\ecor}{\end{corollary}}

\newcommand{\be}{\begin}
\newcommand{\ee}{\end}
\newcommand{\beq}{\begin{equation}}
\newcommand{\eeq}{\end{equation}}
\newcommand{\beqm}{\begin{equation*}}
\newcommand{\eeqm}{\end{equation*}}
\newcommand{\beqn}{\begin{eqnarray}}
\newcommand{\eeqn}{\end{eqnarray}}
\newcommand{\beqnm}{\begin{eqnarray*}}
\newcommand{\eeqnm}{\end{eqnarray*}}
\newcommand{\bea}{\begin{align}}
\newcommand{\eea}{\end{align}}
\newcommand{\beam}{\begin{align*}}
\newcommand{\eeam}{\end{align*}}
\newcommand{\bs}{\begin{subequations}}
\newcommand{\es}{\end{subequations}}

\newcommand{\bei}{\begin{itemize}}
\newcommand{\eei}{\end{itemize}}
\newcommand{\bed}{\begin{description}}
\newcommand{\eed}{\end{description}}
\newcommand{\bee}{\begin{enumerate}}
\newcommand{\eee}{\end{enumerate}}
\newcommand{\bey}{\begin{array}}
\newcommand{\eey}{\end{array}}
\newcommand{\bef}{\begin{figure}}
\newcommand{\eef}{\end{figure}}

\newcommand{\mbf}{\mathbf}

\def\bb[#1]{\mathbb{#1}}
\def\bf[#1]{\mathbf{#1}}
\def\mm[#1]{{\rm #1}}

\def\ff{\frac}

\newcommand{\la}{\label}

\begin{document}

\title{The Quantum Kalman Decomposition: A Gramian Matrix Approach}

\author{Guofeng Zhang\thanks{Department of Applied Mathematics, The Hong Kong Polytechnic University, Hung Hom, Kowloon,  Hong Kong (Guofeng.Zhang@polyu.edu.hk).} 
\and  Jinghao Li\thanks{College of Information Science and Engineering, Northeastern University, Shenyang 110819, China (lijinghao@ise.neu.edu.cn).}
\and  Zhiyuan Dong\thanks{School of Science, Harbin Institute of Technology, Shenzhen, China  (ongzhiyuan@hit.edu.cn).}
\and Ian R. Petersen\thanks{School of Engineering, The Australian National University, Canberra ACT 2601, Australia (i.r.petersen@gmail.com)}
}
\maketitle

\begin{abstract}                         
The Kalman canonical form for quantum linear systems was derived in \cite{ZGPG18}.  The purpose of this  paper is to present an alternative derivation by means of a Gramian matrix approach. Controllability and observability Gramian matrices are defined for linear quantum systems, which are used to characterize various subspaces. Based on these characterizations, real orthogonal and block symplectic coordinate transformation matrices are constructed to transform a given quantum linear system to the Kalman canonical form.  An example is used to illustrate the main results.
\end{abstract}

\textbf{keywords.}
quantum linear control systems,
quantum Kalman canonical form, Gramian matrix

\section{Introduction}
 In recent decades, significant advancements have been made in both theoretical understanding and experimental applications of quantum control. Quantum control plays a pivotal role in various quantum information technologies, such as quantum communication, quantum computation, quantum cryptography, quantum ultra-precision metrology, and nano-electronics. Similar to classical control systems theory, linear quantum systems hold great importance in the field of quantum control. Quantum linear systems are mathematical models that describe the behavior of quantum harmonic oscillators. In this context, ``linear'' refers to the linearity of the Heisenberg equations of motion for quadrature operators in the quantum systems. This linearity often leads to simplifications that facilitate analysis and control of these systems. Consequently, quantum linear systems can be effectively studied using powerful mathematical techniques derived from classical  linear systems theory. A wide range of quantum-mechanical systems can be suitably modeled as quantum linear systems. For instance, quantum optical
systems \cite{WM94,GZ00,WM08,Mabuchi08,WM10,ZJ12,P14,CKS17,NY17,PJU+20,BNC+21}, 
circuit quantum electro-dynamical (circuit QED) systems  \cite{MJP+11,BLS+11,KAK13,BGGW21},
cavity QED systems  \cite{DJ99,SDZ+11,ASD+12}, 
 quantum opto-mechanical
systems \cite{TC10,MHP+11,HM12,DFK+12,MCP+12,NY13,NY14,AKM14,ODP+16,TBCGKP2020,KPS+21,PVB+21,LOW+21},
atomic ensembles \cite{SvHM04,NJP09,NY13,ANP+17,TBCGKP2020}, 
and
quantum memories
 \cite{XDL07,HRG+09,HCH+13,YJ14,NG15}. 

Due to their quantum-mechanical nature, quantum linear systems exhibit several unique control-theoretical properties that do not generally exist in the classical regime. Firstly, stabilizability is equivalent to  detectability (\cite[Section 6.6]{WM10}) and  controllability is equivalent to observability (\cite[Proposition 1]{GZ15}). Secondly, Hurwitz stability implies both controllability and observability (\cite[Theorem 3.1]{ZPL20}). Thirdly, if the system is passive, then Hurwitz stability, controllability, and observability are all equivalent (\cite{GY15},\cite[Lemma 2]{GZ15}). Lastly, the controllable and unobservable  subsystem coexists with the uncontrollable and observable  subsystem \cite{ZGPG18}.

\begin{figure}
  \centering
  \includegraphics[width=0.6\textwidth]{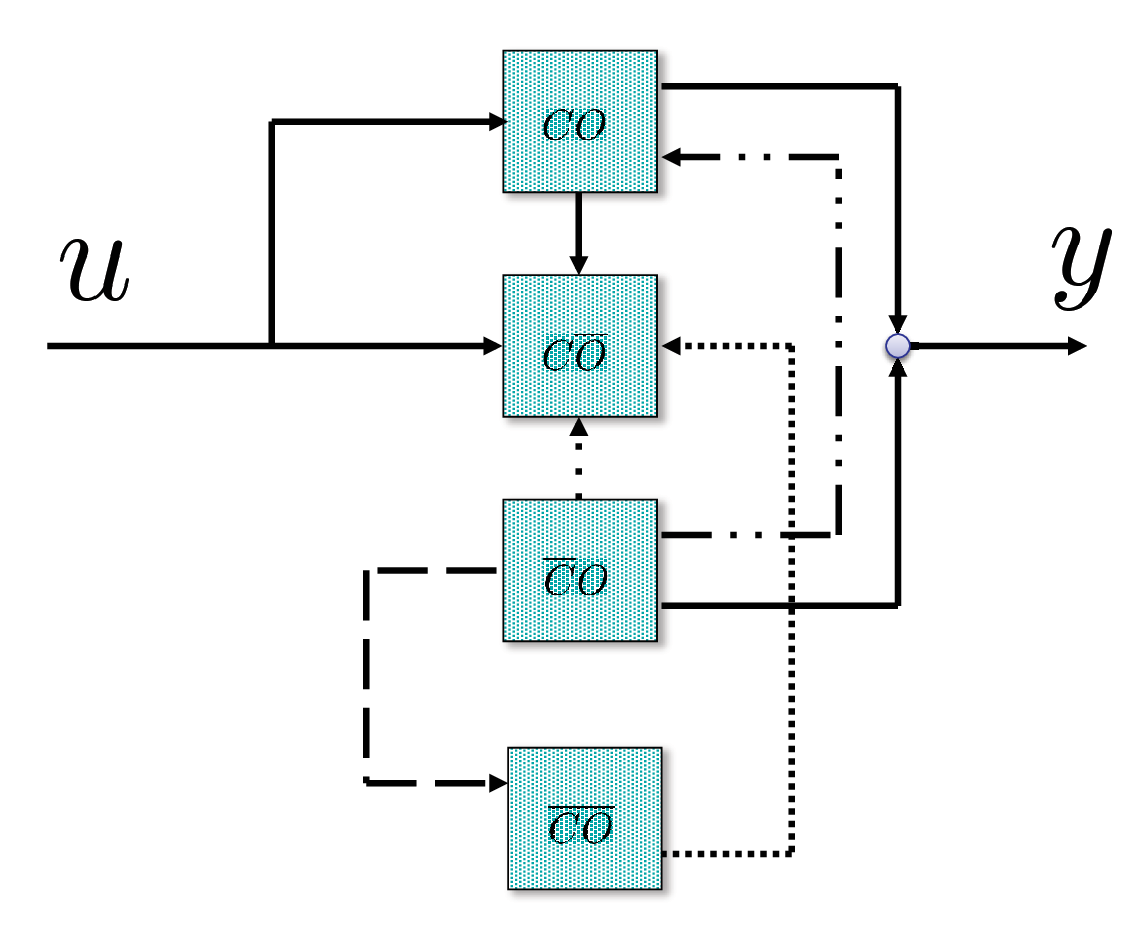}\\
  \caption{The Kalman canonical form of a quantum linear system; see \cite[Fig.~2]{ZGPG18}.} \label{KD}
\end{figure}

The Kalman canonical form, initially proposed for classical linear systems by Kalman in 1963 \cite{Kalman62,Kalman63}, has recently been extended to quantum linear systems \cite{ZGPG18,ZPL20}, where  real orthogonal and block symplectic coordinate transformation matrices are constructed that  transform  a quantum linear system into a new one composed of four possible subsystems: the controllable and observable ($co$) subsystem, the controllable and unobservable ($c\bar{o}$) subsystem, the uncontrollable and observable ($\bar{c}o$) subsystem, and the uncontrollable and unobservable ($\bar{c}\bar{o}$) subsystem. The combination of the $c\bar{o}$ subsystem and the $\bar{c}o$ subsystem is referred to as the ``$h$'' subsystem. As shown in Fig. \ref{KD}, the quantum Kalman canonical form retains the same structure as the classical version but possesses unique properties due to the distinct characteristics of quantum linear systems. Firstly, the controllable and unobservable ($c\bar{o}$) subsystem coexists with the uncontrollable and observable ($\bar{c}o$)  subsystem. Secondly, in the case of a passive system, both the $c\bar{o}$ and $\bar{c}o$ subsystems vanish. Thirdly, the $A$-matrix of the ``$h$'' subsystem and the $A$-matrix of the $\bar{c}\bar{o}$ subsystem are Hamiltonian matrices. In addition to these control-theoretical implications, the quantum Kalman canonical form also provides insights into important physical concepts. For example, the $\bar{c}\bar{o}$ subsystem represents the decoherence-free subsystems (DFSs), and quantum non-demolition (QND) variables reside within the $\bar{c}o$ subsystem, indicating that their temporal evolution  remains unaffected by the input probe or any complementary variables, while still being observable from the output probe. Finally, the determination of quantum back-action evading (BAE) measurement relies solely on the $co$ subsystem. 

The quantum Kalman canonical form is very effective in demonstrating properties of quantum systems. For example, an opto-mechanical system was first theoretically studied in \cite{WC13}, and later on its experimental implementation  was reported in \cite{ODP+16}. This experimental setup successfully demonstrated quantum BAE measurements. In our 2018 paper \cite[Example 5.2]{ZGPG18}, this particular opto-mechanical system was thoroughly examined. By means of the quantum Kalman decomposition,  Equation (83) of \cite{ZGPG18} was obtained which revealed the existence of quantum BAE measurements within this system. Additionally, Equation (83) of \cite{ZGPG18} also predicted the presence of quantum QND variables, denoted as $\boldsymbol{p}_h$ in that equation. Interestingly, a recent experiment conducted in 2021 \cite{KPS+21} demonstrated QND variables precisely corresponding to $\boldsymbol{p}_h$ in Equation (83) of \cite{ZGPG18}. This finding suggests that Equation (83) in \cite{ZGPG18} provides an explanation for both the 2016 quantum BAE experiment \cite{ODP+16} and the 2021 QND experiment \cite{KPS+21}, thus validating the effectiveness of the quantum Kalman canonical form in the study of quantum linear systems theory and experimental quantum physics.

In \cite{ZGPG18}, we start from the annihilation-creation operator representation of quantum linear systems and construct a unitary and block Bogoliubov transformation matrix, then we convert it to a real orthogonal and block symplectic transformation matrix in the quadrature representation. In this paper, we work in the quadrature representation and directly construct real orthogonal and block symplectic transformation matrices.  In particular, we use controllability and observability Gramian matrices   as the main tools in the construction.

The subsequent sections of this paper are organized as follows.  In Section \ref{sec:LQS}, we briefly review quantum linear systems. The observability  and controllability Gramian matrices are presented in Section \ref{sec:gramians}, and used to characterize various subspaces. The construction of the Kalman decomposition for quantum linear systems is given in Section \ref{sec:canonical form}. In Section \ref{sec:numerical} , a computational procedure is given for the construction of coordinate transformation matrices. In Section \ref {sec:example}, an example in the literature is used for demonstration.  Section \ref{sec:con} concludes this paper.

\medskip

\textit{Notation}.
\begin{itemize}
\item $\imath =\sqrt{-1}$ is the imaginary unit. $I_{k}$ is the identity matrix and $0_{k}$ the zero matrix in $\mathbb{C}^{k \times k}$. $\delta_{ij}$ denotes the Kronecker delta;
i.e.,~$I_k=[\delta_{ij}]$. $\delta(t)$ is the Dirac delta function.

\item $\Re(X)$ denotes the real part of $X$ which can be a scalar, vector or matrix, and $\Im(X)$ denotes its imaginary part.


\item $x^{\ast}$ denotes the complex conjugate of a complex number $x$ or
the adjoint of an operator $x$. Clearly. $(xy)^\ast = y^\ast x^\ast$.

\item For a matrix $X=[x_{ij}]$ with  number or operator entries,  $X^{\top}=[x_{ji}]$ is the matrix transpose. Denote $X^{\#}=[x_{ij}^{\ast}]$, and $X^{\dagger}=(X^{\#})^{\top}$. For a vector $x$, we define $\breve{x}\triangleq \bigl[
\begin{smallmatrix}
x \\
x^{\#}
\end{smallmatrix}
\bigr]$.

\item Given two operators $\bf[x]$ and $\bf[y]$, their commutator is defined to be $[\bf[x],\bf[y]] \triangleq \bf[x]\bf[y]-\bf[y]\bf[x]$. Given two \textit{column} vectors of operators $\bf[X]$ and $\bf[Y]$,   their commutator
is defined as
\beq \la{eq:XYT}
[\bf[X],\bf[Y]^\top] \triangleq ([\bf[X]_j,\bf[Y]_k] ) =\bf[X]\bf[Y]^\top- (\bf[Y]\bf[X]^\top)^\top.
\eeq

\item Let $J_{k} \triangleq \mathrm{diag}(I_k,-I_k)$. For a matrix $X\in
\mathbb{C}^{2k\times 2r}$, define its $\flat$-adjoint to be $X^{\flat }
\triangleq J_{r}X^{\dagger}J_{k}$. The $\flat$-adjoint operation enjoys the following  properties:
\beq\la{eq:jun30_flat}
(\alpha A + \beta B)^{\flat}=\alpha^{*}
A^{\flat} + \beta^{*} B^{\flat}, \ \ (AB)^{\flat}=B^{\flat} A^{\flat}, \ \
(A^{\flat})^{\flat}=A,
\eeq
where $\alpha,\beta\in \mathbb{C}$.


\item Given two matrices $U$, $V\in \mathbb{C}^{k\times r}$, define their
 \emph{doubled-up} matrix \cite{GJN10} as  $\Delta
(U,V) \triangleq
\bigl[
\begin{smallmatrix}
U & V \\
V^{\#} & U^{\#}
\end{smallmatrix}
\bigr]$.  The set
of doubled-up matrices is closed under addition, multiplication and $\flat$ adjoint operations.

\item A matrix $T \in \mathbb{C}^{2k\times 2k}$ is called \emph{Bogoliubov}
if it is doubled-up and satisfies $TT^{\flat}=T^{\flat}T=I_{2k}$. The set of Bogoliubov matrices
forms a complex non-compact Lie group known as the Bogoliubov group.

\item Let $\mathbb{J}_{k} \triangleq \bigl[
\begin{smallmatrix}
0_{k} & I_k \\
-I_k & 0_{k}
\end{smallmatrix}
\bigr]$. For a matrix $X\in \mathbb{C}^{2k\times 2r}$, define its $\sharp$-adjoint $X^{\sharp}$ as  $X^{\sharp} \triangleq -\mathbb{J}_{r}X^{\dagger}
\mathbb{J}_{k}$. The $\sharp$-\emph{adjoint} satisfies properties similar to
the usual adjoint, namely
\beq
(\alpha A + \beta B)^{\sharp}=\alpha^{*} A^{\sharp} + \beta^{*}
B^{\sharp}, \ \ (AB)^{\sharp}=B^{\sharp} A^{\sharp},  \ \ (A^{\sharp})^{
\sharp}=A,
\eeq
where $\alpha,\beta\in \mathbb{C}$.

\item A matrix $\mathbb{S} \in \mathbb{C}^{2k\times 2k}$ is called \emph{symplectic},
if $\mathbb{S}\mathbb{S}^{\sharp}=\mathbb{S}^{\sharp}\mathbb{S}=I_{2k}$. Symplectic
matrices forms a complex non-compact group known as the symplectic group.
The subgroup of real symplectic matrices is one-to-one homomorphic to the
Bogoliubov group.

\item A square matrix $M$ is called a Hamiltonian matrix if $(\bb[J]M)^\top = \bb[J]M$. $M^2$ is skew-Hamiltonian as $\bb[J]M^2 = -(\bb[J]M^2)^\top$; See \cite[Section 7.8]{GvL13}.

\item The reduced Planck constant $\hbar$ is set to 1 in this paper.
\end{itemize}

\section{Linear quantum systems} \la{sec:LQS}

\begin{figure}
\includegraphics[width=0.70\textwidth]{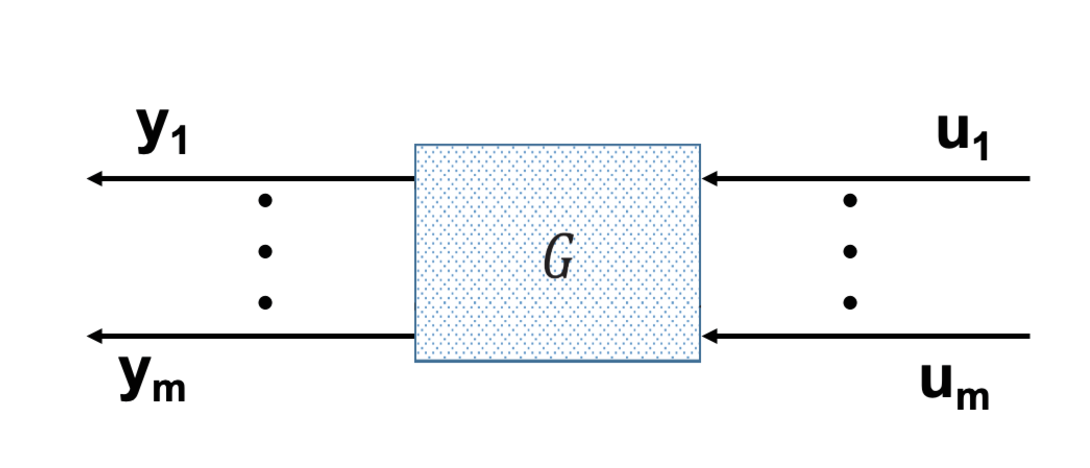}
\centering
\caption{A quantum linear system. Here $G$ consists of  $n$ quantum harmonic oscillators.}
\label{fig:sys}
\end{figure}

The quantum linear system, as shown in Figure \ref{fig:sys}, can be used to model a collection
of $n$ quantum harmonic oscillators driven by $m$ input fields. The quadratures of the $j$-th quantum harmonic oscillator are denoted by $\mbf{q}_j$ and $\mbf{p}_j$ which satisfy the canonical commutation relations $[\mbf{q}_j(t), \ \mbf{p}_j(t)]=\imath$. The $k$-th input field is denoted by $\mbf{u}_k$ whose quadratures $\mbf{q}_{{\rm in},k}$ and $\mbf{p}_{{\rm in},k}$ satisfy the singular commutation relation $[\mbf{q}_{{\rm in},k}(t), \ \mbf{p}_{{\rm in},k}(r)] = \imath\delta(t-r)$.  Similarly,  the $k$-th out field is denoted by $\mbf{y}_k$ whose quadratures $\mbf{q}_{{\rm out},k}$ and $\mbf{p}_{{\rm out},k}$ satisfy the singular commutation relation $[\mbf{q}_{{\rm out},k}(t), \ \mbf{p}_{{\rm out},k}(r)] = \imath\delta(t-r)$. For notational convenience, we denote
\[
\mbf{q}(t) = 
\left[
\bey{c}
\mbf{q}_1(t) \\
\cdots\\
\mbf{q}_n(t) 
\eey
\right], \ 
\mbf{p}(t) = 
\left[
\bey{c}
\mbf{p}_1(t) \\
\cdots\\
\mbf{p}_n(t) 
\eey
\right], \ 
\mbf{q}_{\rm in}(t) = 
\left[
\bey{c}
\mbf{q}_{{\rm in},1}(t) \\
\cdots\\
\mbf{q}_{{\rm in},m}(t)
\eey
\right],
\]
\[
\mbf{p}_{\rm in}(t) = 
\left[
\bey{c}
\mbf{p}_{{\rm in},1}(t) \\
\cdots\\
\mbf{p}_{{\rm in},m}(t)
\eey
\right],\
\mbf{q}_{\rm out}(t) = 
\left[
\bey{c}
\mbf{q}_{{\rm out},1}(t) \\
\cdots\\
\mbf{q}_{{\rm out},m}(t)
\eey
\right], \ 
\mbf{p}_{\rm out}(t) = 
\left[
\bey{c}
\mbf{p}_{{\rm out},1}(t) \\
\cdots\\
\mbf{p}_{{\rm out},m}(t)
\eey
\right].
\]
Denote the system variables, inputs and outputs by
\beq  \label{complex_to_real_trans}
\left[\begin{array}{c}
\mbf{q}(t) \\
\mbf{p}(t)
\end{array}
\right] \equiv  \mbf{x}(t),
\ \ 
\left[\begin{array}{c}
\mbf{q}_{\mathrm{in}}(t) \\
\mbf{p}_{\mathrm{in}}(t)
\end{array}
\right] \equiv  \mbf{u}(t),
\ \ 
\left[\begin{array}{c}
\mbf{q}_{\mathrm{out}}(t) \\
\mbf{p}_{\mathrm{out}}(t)
\end{array}
\right] \equiv \mbf{y}(t),
\eeq
respectively, which satisfy
\beq
[\mbf{x}(t), \ \mbf{x}(t)^\top] = \imath \bb[J]_n, \ [\mbf{u}(t), \ \mbf{u}(r)^\top] =  [\mbf{y}(t), \ \mbf{y}(r)^\top] =\imath \delta(t-r)\bb[J]_m.
\eeq

It is often convenient to describe a quantum system's dynamics in  the $(S,\mbf{L},\mbf{H})$ formalism \cite{GJ09b,GJ09}, as   it offers a powerful modelling framework for analyzing and designing  networked quantum systems; see  e.g., \cite{Mabuchi08,ZJ11,ZJ12,HM12,HM13,WSU+18,KJN18,TBCGKP2020,LDZW22,ZD22,SB23} and references therein. For the quantum linear system in Figure \ref{fig:sys}, $S\in\mathbb{C}^{m\times m}$ is a unitary matrix which can be used to model static devices such as phase shifters and beamsplitters.  The operator $\mbf{L}$ represents the interface between the system $G$ and its inputs, and the operator $\mbf{H}$ describes the Hamiltonian of the system $G$. Mathematically, the coupling operator $\mbf{L}$ and the Hamiltonian $\mbf{H}$ are given by
\beq\la{eq:real_sys_24nov}
\begin{aligned}
\mbf{L} =&\;  \Lambda\mbf{x},
\\
\mbf{H}=&\; \frac{1}{2}\mbf{x}
^\top\mathbb{H} \mbf{x},
\end{aligned}
\eeq
where $\Lambda\in \mathbb{C}^{m\times 2n}$,  and $\bb[H]\in \mathbb{R}^{2n\times 2n}$ is symmetric.

The quantum stochastic differential equations 
(QSDEs) that describe the dynamics of the linear quantum system in Figure \ref{fig:sys} in the real quadrature operator representation are the following:
\beq\la{eq:real_sys}
\begin{aligned}
\dot{\mbf{x}}(t) =&\; \mathbb{A} \mbf{x}(t) + \mathbb{B} \mbf{u}(t),
 \\
\mbf{y}(t) =&\; \mathbb{C} \mbf{x}(t) + \mathbb{D} \mbf{u}(t),
\end{aligned}
\eeq
where the real static system matrices are
\beq  \la{eq:ABCD}
\begin{aligned}
\mathbb{D} =&\;  \left[
\bey{cc}
\Re(S) & -\Im(S) \\
\Im(S)  & \Re(S)
\eey
\right],
\\
\mathbb{C} =&\; 
\sqrt{2}\left[
\bey{c}
\Re(\Lambda)\\
\Im(\Lambda)
\eey
\right],
 \\
\mathbb{B} =&\;   -\mathbb{C}^{\sharp}\mathbb{D},
\\
   \mathbb{A} =&\;  \mathbb{J}_n \mathbb{H} -\frac{1}{2}\mathbb{C}^{\sharp}\mathbb{C}.  
\end{aligned}
\eeq
As the matrix $S$ is unitary,   $\mathbb{D} \mathbb{D}^\sharp = I_{2m}$.
%
%
%
In the real quadrature operator representation \eqref{eq:real_sys} of linear quantum systems, the physical realizability conditions (\cite{JNP08,NJD09,GJN10,SP12,ZJ12,ZJ13}) take  the form
\begin{equation} \la{eq:PR_real}
\mathbb{A} + \mathbb{A}^{\sharp} + \mathbb{B}\mathbb{B}^{\sharp}=0,~\mathbb{B}=-\mathbb{C}^{\sharp}\mathbb{D}^{\sharp}.
\end{equation}

As the scattering matrix $S$ does not affect the coordinate transformation to be performed, in the rest of this paper we assume $S=I_m$.  Consequently,  $\mathbb{D}=I_{2m}$ and $\mathbb{B} = -\mathbb{C}^\sharp$. This can also be understood by regarding $\mathbb{D}\mbf{u}$ as the new input to the system.  More discussions on quantum linear systems can be found in 
\cite{JNP08, NJP09, ZJ11, WM10,NY13,NY17, ZGPG18,ZD22} and references therein.

Note that while we use the
same notation  $\bb[C]$ for the set of complex numbers and the system $C$-matrix in \eqref{eq:real_sys}, the distinction will be self-evident from the context.

\section{Observability  and controllability Gramian matrices} \la{sec:gramians}

In this section, we define observability  and controllability Gramian matrices for the quantum linear system  \eqref{eq:real_sys}, and then use them to characterize various subspaces of $\bb[R]^{2n}$.

For the quantum linear system  \eqref{eq:real_sys}, define the observability and controllability matrices to be
\beq\la{O_G and C_G}
\bb[O]_G \triangleq 
\left[
\bey{c}
\bb[C]\\
\bb[C]\bb[A]
\\
\bb[C]\bb[A]^2
\\
\vdots\\
\bb[C]\bb[A]^{2n-1}
\eey
\right], \ \ \ 
\bb[C]_G \triangleq 
\left[
\bey{ccccc}
\bb[B] & \bb[A]\bb[B] & \bb[A]^2\bb[B] &\cdots& \bb[A]^{2n-1}\bb[B]
\eey
\right],
\eeq
respectively.  We also define two more matrices
\beq\la{O_s and C_s}
\bb[O]_s \triangleq 
\left[
\bey{c}
\bb[C]\\
\bb[C]\bb[J]_n\bb[H]
\\
\bb[C](\bb[J]_n\bb[H])^2
\\
\vdots\\
\bb[C](\bb[J]_n\bb[H])^{2n-1}
\eey
\right], \ \ \ 
\bb[C]_s \triangleq 
\left[
\bey{ccccc}
\bb[B] & (\bb[J]_n\bb[H])\bb[B] & (\bb[J]_n\bb[H])^2\bb[B] &\cdots& (\bb[J]_n\bb[H])^{2n-1}\bb[B]
\eey
\right].
\eeq
The matrices $\bb[O]_s$ and $\bb[C]_s$ are related by
\beq
\left[
\bey{ccccc}
\bb[J]_m &&&&\\
& -\bb[J]_m &&&\\
& &\ddots &&\\
&&&\bb[J]_m &\\
&&&&-\bb[J]_m
\eey
\right]\bb[C]_s^\top \bb[J]_n = \bb[O]_s.
\eeq
Consequently,
\beq\la{eq:10_nov_temp1}
\begin{aligned}
{\rm Ker}(\bb[O]_s) =& \;  {\rm Ker}(\mathbb{C}_s^\top \bb[J]_n),
\\
{\rm Ker}(\bb[O]_s \bb[J]_n) =&\; {\rm Ker}(\mathbb{C}_s^\top ) = {\rm Im}(\bb[C]_s)^\perp.
\end{aligned}
\eeq
It can easily checked that 
\beq \la{eq:equiv_9_Nov}
{\rm Im}(\bb[C]_G) =  {\rm Im}(\bb[C]_s), \ 
{\rm Ker}(\bb[O]_G) =  {\rm Ker}(\bb[O]_s).
\eeq
Hence, we use $\bb[O]_s$ and $\bb[C]_s$, instead of  $\bb[O]_G$ and $\bb[C]_G$, in the following discussions as they appear simpler.

\bmrk
Notice that the matrix $\bb[O]_s$ defined in Eq. \eqref{O_s and C_s} is the real-domain counterpart of the matrix $\mbf{O}_s$ defined in \cite[Eq. (7)]{GZ15} in the complex domain.
\emrk

Similar to the classical case, see e.g., \cite{JH04} and \cite[Chapter 9]{Rugh96},  define the observability Gramian matrix $W_o(t,s)$ and controllability Gramian matrix $W_c(t,s)$  to be
\bs\la{eq:gramian}
\begin{align}
W_o(t,s) \triangleq  \; 
\int_t^s e^{(\bb[J]_n\bb[H])^\top(\tau-t)}\bb[C]^\top \bb[C] e^{(\bb[J]_n\bb[H])(\tau-t)}d\tau,  \ \ \forall -\infty <t<s<\infty,
\la{W_o}\\
W_c(t,s) \triangleq  \; 
\int_t^s e^{(\bb[J]_n\bb[H])(t-\tau)}\bb[B]\bb[B]^\top e^{(\bb[J]_n\bb[H])^\top(t-\tau)}d\tau ,   \ \ \forall -\infty <t<s<\infty,
\la{W_c}
\end{align}
\es
respectively. Clearly, both $W_o(t,s)$ and $W_c(t,s)$ are real and positive semi-definite matrices. 

The following result reveals a nice relation between the  observability and  controllability  Gramians  which is unique to quantum linear systems,
\blem \la{lem:W_o_Ws}
The observability Gramian $W_o(t,s)$ and  controllability  Gramian $W_c(t,s)$ are related by
\beq \la{eq:Wo_Wc}
W_o(t,s) = W_c(t,s)^\sharp.
\eeq
\elem

\textbf{Proof.} 
 Noticing that $\bb[J]_n^\top = -\bb[J]_n$ and  $\bb[J]_n^\top e^{\bb[J]_n \bb[H]}\bb[J]_n = e^{ \bb[H]\bb[J]_n}$, we have
\beqn
W_c(t,s)^\sharp &=& \bb[J]_n^\top W_c(t,s) \bb[J]_n 
\nonumber\\
&=&
\int_t^s \bb[J]_n^\top e^{(\bb[J]_n\bb[H])(t-\tau)}\bb[B]\bb[B]^\top e^{(\bb[J]_n\bb[H])^\top(t-\tau)} \bb[J]_n d\tau
\nonumber\\
&=&
\int_t^s \left(\bb[J]_n^\top e^{(\bb[J]_n\bb[H])(t-\tau)}\bb[J]_n\right)\left(\bb[J]_n^\top \bb[B]\bb[J]_m\right) \left(\bb[J]_n^\top \bb[B]\bb[J]_m\right)^\top  \left( \bb[J]_n^\top e^{(\bb[J]_n\bb[H])(t-\tau)} \bb[J]_n \right)^\top d\tau
\nonumber\\
&=&
\int_t^s \left( e^{ (\bb[H]\bb[J]_n)(t-\tau)}  \right) \left(-\bb[C]^\top \right)\left(-\bb[C] \right) \left( e^{ (\bb[H]\bb[J]_n)(t-\tau)}\right)^\top
\nonumber\\
&=& 
\int_t^s e^{(\bb[J]_n\bb[H])^\top(\tau-t)}\bb[C]^\top \bb[C] e^{(\bb[J]_n\bb[H])(\tau-t)}d\tau
\nonumber\\
&=&
W_o(t,s).
\la{eq:25Nov_Wo_Ws}
\eeqn
This completes the proof. 
$\Box$

In the following, We characterize various subspaces of $\bb[R]^{2n}$  in terms of the observability Gramian matrix $W_o(t,s)$ and controllability Gramian matrix $W_c(t,s)$.


We start from the following result.
\blem \la{lem:O}
We have
\beq\la{eq:Ker_Os_Wo}
{\rm Ker}(\bb[O]_s) = {\rm Ker}(W_o(t,s)), \ \ \forall -\infty <t<s<\infty,
\eeq
and 
\beq\la{eq:Im_Os_Wo}
{\rm Im}(\bb[O]_s^\top) = {\rm Im}(W_o(t,s)), \ \ \forall -\infty <t<s<\infty.
\eeq
\elem
\textbf{Proof.} 
For each $x\in {\rm Ker}(\bb[O]_s)$, by Eq. \eqref{O_s and C_s} we have $\bb[C](\bb[J]_n\bb[H])^kx =0$ for all $k=0,1,\ldots, 2n-1$. Hence, according to the Cayley-Hamilton theorem,  $\bb[C]e^{(\bb[J]_n\bb[H])t}x =0$ for all $t\in\bb[R]$, which means that $x\in {\rm Ker}(W_o(t,s))$ for all $-\infty <t<s<\infty$. Hence, ${\rm Ker}(\bb[O]_s) \subset {\rm Ker}(W_o(t,s))$ for all  $-\infty <t<s<\infty$. On the other hand, suppose $x\in {\rm Ker}(W_o(t,s))$ for all  $-\infty <t<s<\infty$. Then $\int_t^s\|\bb[C] e^{(\bb[J]_n\bb[H])(\tau-t)}x\|^2 d\tau =0$, which means that $\bb[C] e^{(\bb[J]_n\bb[H])(\tau-t)}x\equiv 0$ for all  $t\leq \tau \leq s$. In other words, $\bb[C] e^{(\bb[J]_n\bb[H])\tau} x\equiv 0$ for all $\tau \in [0, s-t]$. As the derivative $ \ff{{\rm d} e^{(\bb[J]_n\bb[H])\tau}}{{\rm d} \tau} = \bb[J]_n\bb[H] e^{(\bb[J]_n\bb[H])\tau}$, we have $\bb[C](\bb[J]_n\bb[H])^k  e^{(\bb[J]_n\bb[H])\tau} x\equiv 0$ for all $\tau \in [0,s-t]$  and for all $k=0,1,\ldots$. Setting $\tau=0$ yields $\bb[C](\bb[J]_n\bb[H])^k x\equiv 0$  for all $k=0,1,\ldots$, which means that $x\in {\rm Ker}(\bb[O]_s)$.  Thus,  $ {\rm Ker}(W_o(t,s))\subset {\rm Ker}(\bb[O]_s)$. Consequently, Eq. \eqref{eq:Ker_Os_Wo} holds. Eq. \eqref{eq:Im_Os_Wo} immediately follows Eq. \eqref{eq:Ker_Os_Wo}. 
$\Box$

\bmrk\la{rem:constant_rank}
Lemma \ref{lem:O} indicates that the rank of $W_o(t,s)$ and $W_c(t,s)$ is constant  as long as $t<s$.
\emrk

By Eqs. \eqref{eq:10_nov_temp1}, \eqref{eq:Wo_Wc}, and Lemma \ref{lem:O} above, we have the following result.
\blem \la{lem:C}
We have 
\beq \la{Im_Cg_Wc}
{\rm Im}(\bb[C]_s) = {\rm Im}(W_c(t,s)), \ \ \forall -\infty <t<s<\infty
\eeq
and 
\beq
{\rm Ker}(\bb[C]_s^\top) = {\rm Ker}(W_c(t,s)), \ \ \forall -\infty <t<s<\infty.
\eeq
\elem

Based on Lemmas \ref{lem:W_o_Ws}-\ref{lem:C}, we have  the following result.
\bthm 
The space $\mathbb{R}^{2n}$ can be divided as
\beq \la{eq:25_nov_space}
\mathbb{R}^{2n}=R_{c\bar{o}}\oplus
R_{co}\oplus R_{\bar{c}\bar{o}}\oplus R_{\bar{c}o},
\eeq
where 
\beq\la{eq:gwge_22nov}
\begin{aligned}
  R_{c\bar{o}} \triangleq &\;   {\rm Im}(\bb[J]_n W_o(t,s)) \cap {\rm Ker}(W_o(t,s)),
 \\
   R_{co} \triangleq &\;   {\rm Im}(\bb[J]_n W_o(t,s)) \cap {\rm Im}(W_o(t,s)),
 \\
  R_{\bar{c}\bar{o}} \triangleq &\;   {\rm Ker}( W_o(t,s)\bb[J]_n) \cap {\rm Ker}(W_o(t,s)),
 \\
   R_{\bar{c}o} \triangleq &\;   {\rm Ker}( W_o(t,s)\bb[J]_n) \cap {\rm Im}(W_o(t,s)).
\end{aligned}
\eeq
\ethm

\textbf{Proof.} 
 By Eqs. \eqref{eq:Wo_Wc} and \eqref{eq:Ker_Os_Wo},
\beq\la{eq:9nov_yerh}
{\rm Ker}(\bb[C]_s^\top) =  {\rm Ker}(\bb[J]_n W_o(t,s)\bb[J]_n) = {\rm Ker}( W_o(t,s)\bb[J]_n) = {\rm Ker}( \bb[O]_s\bb[J]_n). 
\eeq

Similarly, by Eqs. \eqref{eq:Wo_Wc} and \eqref{Im_Cg_Wc}, 
\beq
{\rm Im}(\bb[C]_s) ={\rm Im}(\bb[J]_n W_o(t,s)) =  \bb[J]_n{\rm Im}(W_o(t,s)).
\eeq
As a result,
\beq\la{eq:25_nov_temp1}
\be{aligned}
{\rm Im}(\bb[C]_s)\cap {\rm Ker}(\bb[O]_s) 
=&
{\rm Im}(\bb[J]_n W_o(t,s))\cap  {\rm Ker}(W_o(t,s)) = R_{c\bar{o}},
\\
{\rm Im}(\bb[C]_s)\cap {\rm Im}(\bb[O]_s^\top)
=&
{\rm Im}(\bb[J]_n W_o(t,s))\cap {\rm Im}(W_o(t,s)) = R_{co},
\\
{\rm Ker}(\bb[C]_s^\top)  \cap {\rm Ker}(\bb[O]_s) 
=&
{\rm Ker}( W_o(t,s)\bb[J]_n)  \cap  {\rm Ker}(W_o(t,s)) = R_{\bar{c}\bar{o}},
\\
{\rm Ker}(\bb[C]_s^\top)  \cap {\rm Im}(\bb[O]_s^\top)
=&
{\rm Ker}( W_o(t,s)\bb[J]_n)  \cap {\rm Im}(W_o(t,s)) = R_{\bar{c}o}
\ee{aligned}
\eeq
Hence, Eq. \eqref{eq:25_nov_space} follows Eq. \eqref{eq:25_nov_temp1}. 
$\Box$

Clearly, Eq. \eqref{eq:25_nov_space} can be re-written as
\beq \la{eq:gwge}
\begin{aligned}
  R_{c\bar{o}} \triangleq &\;  {\rm Ker}( W_o(t,s)\bb[J]_n)^\perp \cap {\rm Ker}(W_o(t,s)),
  \\
   R_{co} \triangleq &\;   {\rm Ker}( W_o(t,s)\bb[J]_n)^\perp \cap {\rm Ker}(W_o(t,s))^\perp,
  \\
  R_{\bar{c}\bar{o}} \triangleq &\;   {\rm Ker}( W_o(t,s)\bb[J]_n) \cap {\rm Ker}(W_o(t,s)),
  \\
   R_{\bar{c}o} \triangleq &\;   {\rm Ker}( W_o(t,s)\bb[J]_n) \cap {\rm Ker}(W_o(t,s))^\perp.
\end{aligned}
\eeq
By Eq. \eqref{eq:gwge_22nov} or Eq. \eqref{eq:gwge}, we have the following result.
\bcor
\beq \la{eq:J_subspace}
R_{co} = \bb[J]_n R_{co}, \ \ R_{\bar{c}\bar{o}} = \bb[J]_n R_{\bar{c}\bar{o}}, \ \ R_{c\bar{o}} =\bb[J]_n R_{\bar{c}o}.
\eeq
\ecor

\bmrk
Notice that Eq. \eqref{eq:J_subspace} is the real-domain counterpart of \cite[Eq. (27)]{ZGPG18} in the complex domain.
\emrk

\section{Kalman decomposition for quantum linear systems} \la{sec:canonical form}

In this section, we construct the coordinate transformation matrix that transform the quantum linear system \eqref{eq:real_sys} to the Kalman canonical form as shown in Fig. \ref{KD}.

\subsection{Subspace $R_{co}$}

If the unit vector $\left[\bey{c} e_1\\ f_1 \eey\right] \in R_{co}$, then by Eq. \eqref{eq:J_subspace}, the two vectors
\[
\left[\bey{c} 
e_1+ f_1\\
-(e_1- f_1)
\eey\right], \left[\bey{c} 
e_1- f_1\\
e_1+ f_1
\eey\right] \in R_{co},
\]
and they are orthogonal to each other too. 
Choose another unit vector $\left[\bey{c} e_2\\ f_2 \eey\right] \in R_{co}$ which is orthogonal to both 
\[
\left[\bey{c} 
e_1+ f_1\\
-(e_1- f_1)
\eey\right], \left[\bey{c} 
e_1- f_1\\
e_1+ f_1
\eey\right].
\]
Then it is straightforward to show that all
the vectors
\beq\la{eq:17Novtgge}
\left[\bey{c} 
e_1+ f_1\\
-(e_1- f_1)
\eey\right], \left[\bey{c} 
e_1- f_1\\
e_1+ f_1
\eey\right], \left[\bey{c} 
e_2+ f_2\\
-(e_2- f_2)
\eey\right], \left[\bey{c} 
e_2- f_2\\
e_2+ f_2
\eey\right]\in R_{co}
\eeq
are orthogonal to each other. Thus, the dimension of $R_{co}$ must be even, which is denoted by $2n_1$ for some non-negative integer $n_1$. Repeat the above procedure to get $2n_1$ orthogonal vectors
\[
\left[\bey{c} 
e_1+ f_1\\
-(e_1- f_1)
\eey\right], \left[\bey{c} 
e_1- f_1\\
e_1+ f_1
\eey\right], \left[\bey{c} 
e_2+ f_2\\
-(e_2- f_2)
\eey\right], \left[\bey{c} 
e_2- f_2\\
e_2+ f_2
\eey\right], \cdots, \left[\bey{c} 
e_{n_1}+ f_{n_1}\\
-(e_{n_1}- f_{n_1})
\eey\right], \left[\bey{c} 
e_{n_1}- f_{n_1}\\
e_{n_1}+ f_{n_1}
\eey\right] \in R_{co}
\]
Define a matrix
\beq \la{eq:T_co}
T_{co} \triangleq \ff1{\sqrt{2}}
\left[
\bey{cccccc}
e_1+ f_1 & \cdots & e_{n_1}+ f_{n_1} & e_1- f_1 & \cdots &  e_{n_1}- f_{n_1} \\
-(e_1- f_1) & \cdots  &  -(e_{n_1}- f_{n_1}) & e_1+ f_1 &\cdots & e_{n_1}+ f_{n_1}
\eey
\right].
\eeq
The above construction guarantees that  $T_{co}$ is real orthogonal. Moreover,
as
\beq \la{eq:sym_n1}
T_{co}^\top \bb[J]_{n} T_{co} =  \bb[J]_{n_1},
\eeq
 $T_{co}$ is also symplectic. Define system variables
\beq
\bf[x]_{co} \triangleq T_{co}^\top \bf[x].
\eeq
We have 
\beq
[\bf[x]_{co}, \ \bf[x]_{co}^\top] = \imath \bb[J]_{n_1}.
\eeq
In other words, the coordinate transformation $T_{co}$ preserves the canonical commutation relations.

\subsection{Subspace $R_{\bar{c}\bar{o}}$}

Similarly, let the dimension of the subspace $R_{\bar{c}\bar{o}}$ be $2n_2$ for some non-negative integer $n_2$. One can construct a real orthogonal and symplectic $T_{\bar{c}\bar{o}}$ of the form \eqref{eq:T_co}. 
Define system variables
\beq
\bf[x]_{\bar{c}\bar{o}} \triangleq T_{\bar{c}\bar{o}}^\top \bf[x].
\eeq
We have
\beq
[\bf[x]_{\bar{c}\bar{o}}, \ \bf[x]_{\bar{c}\bar{o}}^\top] = \imath \bb[J]_{n_2},
\eeq
and
\beq\la{eq:sym_n2}
T_{\bar{c}\bar{o}}^\top \bb[J]_n T_{\bar{c}\bar{o}} = \bb[J]_{n_2}.
\eeq

\subsection{The ``h'' subspace}

As introduced in the {\it Notation} part, given two matrices $M$, $N\in \mathbb{C}^{k\times \ell}$, the corresponding doubled-up matrix is 
\beq\la{eq:doubled_up}
\Delta(M,N) = \left[
\bey{cc}
M & N\\
N^\# & M^\#
\eey
\right].
\eeq
One may define another operation as
\beq \la{eq:tuild_doubped_up}
\tilde{\Delta}(M,N) \triangleq V_k \Delta(M,N) V_ \ell^\dag
=\left[
\bey{cc}
\Re(M+N) & -\Im(M-N)\\
\Im(M+N) & \Re(M-N) 
\eey
\right],
\eeq 
where  the unitary matrix $V_k$ is given by
\beq
V_k = \ff1{\sqrt{2}}
\left[
\bey{cc}
I_k & I_k\\
-\imath I_k & \imath I_k
\eey
\right].
\eeq 
Given a complex vector $a\in \mathbb{C}^{2 \ell}$, by Eq. \eqref{eq:tuild_doubped_up} we have
\beq
\tilde{\Delta}(M,N) \left[
\bey{c}
\Re(a)\\
\Im(a)
\eey
\right]
=
\left[
\bey{c}
\Re(Ma+Na^\#)\\
\Im(Ma+Na^\#)
\eey
\right].
\eeq
From
\beq
\Delta(M_1,N_1)\Delta(M_2,N_2) = \Delta(M_1M_2+N_1N_2^\#, M_1N_2+N_1 M_2^\#)
\eeq
we know that
\beq
\tilde{\Delta}(M_1,N_1)\tilde{\Delta}(M_2,N_2) =
\tilde{\Delta}(M_1M_2+N_1N_2^\#, M_1N_2+N_1 M_2^\#).
\eeq
Moreover,  from 
\beq
(\Delta(M_1,N_1)\Delta(M_2,N_2) )^\flat =\Delta(M_2,N_2) ^\flat \Delta(M_1,N_1) ^\flat ,
\eeq
we have
\beq
(\tilde{\Delta}(M_1,N_1)\tilde{\Delta}((M_2,N_2))^\sharp 
=
\tilde{\Delta}((M_2,N_2))^\sharp \tilde{\Delta}(M_1,N_1)^\sharp
\eeq 
In fact, the set
of  matrices of the form \eqref{eq:tuild_doubped_up} is closed under addition, multiplication and $\sharp$-adjoint operation.

\bmrk
All the system matrices in Eq. \eqref{eq:ABCD}
 are in the form of $\tilde{\Delta}(M,N)$ defined in Eq. \eqref{eq:tuild_doubped_up}.
\emrk

In \cite{ZGPG18} the doubled-up matrices  of the form \eqref{eq:doubled_up} play a crucial rule in the construction of special orthonormal bases for the subspaces $R_{c\bar{o}}$ and $R_{\bar{c}o}$; See \cite[Lemmas 4.4-4.7]{ZGPG18} for details. Notice that the set of matrices of the form \eqref{eq:tuild_doubped_up} is the counterpart of the set of doubled-up matrices. Hence one may attempt to construct special orthonormal bases for the subspaces $R_{c\bar{o}}$ and $R_{\bar{c}o}$ by means of the set of matrices of the form \eqref{eq:tuild_doubped_up} by   adopting similar tricks as those in \cite{ZGPG18}. However, in this paper we will use a   simpler method which relies on a special property of the the subspace $R_{c\bar{o}}$ to be given in Proposition \ref{prop_cobar}.

Let the dimension of the subspace $R_{c\bar{o}}$ be $n_3$ for some non-negative integer $n_3$. Then let 
\beq\la{eq:T_cbaro}
T_{c\bar{o}} \triangleq
\left[
\bey{cccc}
g_1 & g_2 & \cdots & g_{n_3} \\
h_1 & h_2 & \cdots & h_{n_3} 
\eey
\right] \in R_{c\bar{o}}
\eeq
be a real orthonormal matrix.

The orthonormal vectors in the matrix $T_{c\bar{o}}$ defined  in Eq. \eqref{eq:T_cbaro} enjoy the following property.
\bprop \la{prop_cobar}
\beq\la{eq:Cbaro_perp_barco}
\left[
\bey{c}
g_j \\
h_j 
\eey
\right] \perp \bb[J]_n \left[
\bey{c}
g_k \\
h_k 
\eey
\right],  \ \ \forall j, k= 1,2, \ldots n_3.
\eeq
\eprop

\textbf{Proof.} 
Given any column vector $\left[
\bey{c}
g_j \\
h_j 
\eey
\right]$ of $T_{c\bar{o}}$, by
Eqs. \eqref{eq:gwge} we get
\[
\left[
\bey{c}
g_j \\
h_j 
\eey
\right]  \perp R_{\bar{c}o} .
\]
On the other hand, from Eq. \eqref{eq:J_subspace} we know that $\bb[J]_n \left[
\bey{c}
g_k \\
h_k 
\eey
\right] \in R_{\bar{c}o}$ for any column vector $\left[
\bey{c}
g_k \\
h_k 
\eey
\right] $ of $T_{c\bar{o}}$.  
As $R_{c\bar{o}} \perp R_{\bar{c}o}$, Eq. \eqref{eq:Cbaro_perp_barco} holds.  
$\Box$

By means of $T_{c\bar{o}}$ defined  in Eq. \eqref{eq:T_cbaro},  we  define another real orthonormal matrix 
\beq\la{eq:16nov_T_barco}
T_{\bar{c}o} \triangleq \bb[J]_n^\top T_{c\bar{o}} \in \mathbb{R}_{\bar{c}o}.
\eeq
Then define system variables
\beq
\bf[x]_{c\bar{o}} \triangleq T_{c\bar{o}}^\top \bf[x], \ \ 
\bf[x]_{\bar{c}o} \triangleq T_{\bar{c}o}^\top \bf[x].
\eeq
It is easy to show that 
\beq
[\bf[x]_{c\bar{o}}, \ \ \bf[x]_{\bar{c}o}^\top]
=
T_{c\bar{o}}^\top[\bf[x], \ \ \bf[x]^\top]T_{\bar{c}o}
=
\imath T_{c\bar{o}}^\top \bb[J]_{n}\bb[J]_n^\top T_{c\bar{o}}
=\imath I_{n_3}.
\eeq

\subsection{Quantum Kalman decomposition}

We are ready to transform the linear quantum system  \eqref{eq:real_sys} into its Kalman canonical form. Define
\beq \la{eq:T_final}
T=
\left[
\bey{cccc}
T_{c\bar{o}} & T_{co}& T_{\bar{c}\bar{o}}&T_{\bar{c}o}
\eey
\right] .
\eeq

\bthm
We have
\beqn
T^\top W_c(t,s)T
&=&\left[
\begin{array}{cccc}
W^c_{11}(t,s) & W^c_{12}(t,s) & 0_{n_3\times 2n_2} & 0_{n_3\times n_3}\\
W^c_{12}(t,s)^\top & W^c_{22}(t,s) & 0_{2n_1\times 2n_2} & 0_{2n_1\times n_3}\\
0_{2n_2\times n_3} & 0_{2n_2\times 2n_1} & 0_{2n_2\times 2n_2} & 0_{2n_2\times n_3}\\
0_{n_3\times n_3} & 0_{n_3\times 2n_1} & 0_{n_3\times 2n_2} & 0_{n_3\times n_3}
\end{array}
\right],
\la{eq:W_c_kalman}
\\
T^\top W_o(t,s)T
&=& \left[
\begin{array}{cccc}
0_{n_3\times n_3} & 0_{n_3\times 2n_1} & 0_{n_3\times 2n_2} & 0_{n_3\times n_3}\\
0_{2n_1\times n_3} & W^o_{22}(t,s) & 0_{2n_1\times 2n_2} & W^o_{24}(t,s)^\top\\
0_{2n_2\times n_3} & 0_{2n_2\times 2n_1} & 0_{2n_2\times 2n_2} & 0_{2n_2\times n_3}\\
0_{n_3\times n_3} & W^o_{24}(t,s) & 0_{n_3\times 2n_2} & W^o_{44}(t,s)
\end{array}
\right].
\la{eq:W_o_kalman}
\eeqn
Moreover, the transformation $\bar{\mbf{x}} = T^\top \mbf{x}$ transforms system \eqref{eq:real_sys}  into 
\begin{equation} \label{real_Kalman_ss}
\begin{split}
\boldsymbol{\dot{\bar{x}}}(t) 
=&\;
\bar{A}\boldsymbol{\bar{x}}(t)  +\bar{B}\boldsymbol{u}(t),
\\
\boldsymbol{y}(t)
=&\;
 \bar{C}\boldsymbol{\bar{x}}(t) + \boldsymbol{u}(t),
\end{split}
\end{equation}
where 
\beq \la{eq:kalman_ABC}
\bar{A} = T^\top \bb[A] T =
\left[
\bey{cccc}
\bb[A]_{h}^{11} & \bb[A]_{12} & \bb[A]_{13} & \bb[A]_{h}^{12} \\ 
0 & \bb[A]_{co} & 0 & \bb[A]_{24} \\ 
0 & 0 & \bb[A]_{\bar{c}\bar{o}} & \bb[A]_{34} \\ 
0 & 0 & 0 & \bb[A]_{h}^{22}
\eey
\right], \ 
\bar{B} = T^\top \bb[B] =
\left[
\bey{c}
\bb[B]_{h} \\ 
\bb[B]_{co} \\ 
0 \\ 
0
\eey
\right], \ 
\bar{\bb[C]} =
 \bb[C]T =
 \left[
\bey{cccc}
  0 & \bb[C]_{co} & 0 & \bb[C]_{h}
\eey
\right].
\eeq
\ethm

\textbf{proof.}  Firstly, by Eqs. \eqref{Im_Cg_Wc} and \eqref{eq:9nov_yerh},
\beq \la{wc_perp}
{\rm Im}(W_c(t,s)) \perp  {\rm Ker}(W_o(t,s)\bb[J]_n).
\eeq 
According to Eq. \eqref{wc_perp},
\beq
W_c(t,s) \perp [T_{\bar{c}\bar{o}} \ \  T_{\bar{c}o}].
\eeq
This implies Eq. \eqref{eq:W_c_kalman}.  Secondly, from
\[
{\rm Im}(W_o(t,s)) \perp  {\rm Ker}(W_o(t,s)),
\]
we have
\beq
W_o(t,s) \perp [T_{c\bar{o}} \ \ T_{\bar{c}\bar{o}}].
\eeq
This implies Eq. \eqref{eq:W_o_kalman}. Finally, by means of the well-known invariance
properties of linear systems; e.g., see \cite[Chapter 2]{Kimura96} and \cite[Chapter 6]{CF03}: 
\begin{equation}
\bb[A]R_{c\bar{o}}\subset R_{c\bar{o}},~\bb[A]R_{co}\subset R_{c%
\bar{o}} \oplus R_{co}, ~\bb[A]R_{\bar{c}\bar{o}}\subset R_{c\bar{o}}
\oplus R_{\bar{c}\bar{o}}  \label{subset_a}
\end{equation}
and 
\begin{eqnarray}
&& \mathrm{Im}(\bb[B]) \subset \mathrm{Im}(C_s)=R_{c\bar{o}} \oplus
R_{co},  \notag \\
&& \mathrm{Ker}(O_s) = R_{c\bar{o}} \oplus R_{\bar{c}\bar{o}} \subset 
\mathrm{Ker}(\bb[C]),  \label{subset_b}
\end{eqnarray}
we have  Eq. \eqref{real_Kalman_ss}. 
$\Box$

Define
\beq\la{eq:Th_22_Nov}
T_h \triangleq [T_{c\bar{o}} \ \ T_{\bar{c}o}].
\eeq
We have
\beq\la{eq:sym_n3}
T_{h}^\top \bb[J]_n T_{h} = \bb[J]_{n_3}.
\eeq
Then we can rewrite the matrix $T$ in Eq. \eqref{eq:T_final} as 
\beq\la{eq:T_final_2}
\tilde{T}=\left[\begin{array}{c|c|c}
 T_h & T_{co} & T_{\bar{c}\bar{o}}
\end{array}\right].
\eeq
Clearly, $T$ is real orthogonal. Furthermore,  according to Eqs. \eqref{eq:sym_n1}, \eqref{eq:sym_n2}, and \eqref{eq:sym_n3},  we have
\beq
\tilde{T}^\top\bb[J]_n \tilde{T} = \left[
\bey{ccc}
\bb[J]_{n_3} & &\\
& \bb[J]_{n_1} & \\
& & \bb[J]_{n_2}
\eey
\right].
\eeq
In other words, $\tilde{T}$ is  block-wise  symplectic.

\section{A procedure for computing the coordinate transformation matrix}\la{sec:numerical}

In the preceding section,  we showed how to obtain the coordinate transformation matrix $T$  starting from orthonormal bases of the subspaces. In this section, we propose methods for finding orthonormal bases for these subspaces, thus completing the whole procedure of computing the coordinate transformation matrix.

We apply  the singular value decomposition (SVD) to the observability Gramian matrix $W_o(t,s)$. Since $W_o(t,s)$ is real positive semi-definite, there exists an orthogonal matrix $U=\left[U_1\ U_2\right]\in\mathbb{R}^{2n\times 2n}$ such that
\begin{equation}\label{wo_decomp}
    W_o(t,s)=U\Sigma U^\top=U\left[\begin{array}{cc}
    \Sigma_1 & 0 \\
    0 & 0_{(2n-r)\times(2n-r)}
\end{array}\right]U^\top=U_1\Sigma_1U_1^\top,
\end{equation}
where $\Sigma_1\in\mathbb{R}^{r\times r}$ is a  diagonal matrix with positive diagonal entries. Therefore, $\mathrm{rank}(U_1)=\mathrm{rank}(W_o(t,s))=r=2n_1+n_3$ and  $\mathrm{rank}(U_2)=2n-r$. As mentioned in Remark \ref{rem:constant_rank}, $r$ is constant as long as $t<s$. Thus, without loss of generality, in Eq. \eqref{wo_decomp} above we implicitly assumed $t$ and $s$ are two fixed constant satisfying $t<s$. Thus the RHS of Eq. \eqref{wo_decomp} does  not depend on $t$ and $s$.

We characterize the subspaces in Eq. \eqref{eq:gwge_22nov} by means of orthogonal matrices $U_1$ and $U_2$ given above.

\blem The subspaces can be expressed as
\beq\la{eq:gwgegewe_22nov}
\begin{aligned}
  R_{c\bar{o}} = &\;   \mathrm{Im}(\mathbb{J}U_1) \cap \mathrm{Im}(U_2),
  \\
   R_{co} = &\;    \mathrm{Im}(\mathbb{J}U_1) \cap\mathrm{Im}(U_1),
  \\
  R_{\bar{c}\bar{o}} = &\;   \mathrm{Im}(\mathbb{J}U_2) \cap \mathrm{Im}(U_2),
  \\
   R_{\bar{c}o} = &\;   \mathrm{Im}(\mathbb{J}U_2) \cap \mathrm{Im}(U_1),
\end{aligned}
\eeq
respectively.
\elem

\textbf{Proof.} 
 From Eqs. \eqref{eq:Wo_Wc} and \eqref{wo_decomp}, it can be easily seen  that the controllability Gramian matrix  $W_c(t,s)$ has a SVD of the form
\begin{equation}\label{wc_decomp}
W_c(t,s)=(\mathbb{J}U)\Sigma(\mathbb{J}U)^\top=(\mathbb{J}U_1)\Sigma_1(\mathbb{J}U_1)^\top.
\end{equation}
%
By the properties of the SVD, we have
\begin{equation}\label{wandu}
    \mathrm{Ker}(W_o(t,s))=\mathrm{Im}(U_2),\ 
\mathrm{Im}(W_o(t,s))=\mathrm{Im}(U_1),\ 
\mathrm{Im}(W_c(t,s))=\mathrm{Im}(\mathbb{J}U_1),\
\mathrm{Ker}(W_c(t,s))=\mathrm{Im}(\mathbb{J}U_2).
\end{equation}
Using the relationship in Eq. \eqref{wandu}, the subspaces in Eq. \eqref{eq:gwge_22nov} can be rewritten in terms of $U_1$ and $U_2$ as those in Eq. \eqref{eq:gwgegewe_22nov}. 
$\Box$

In the following, we express these four subspaces in an alternative way for ease of numerical computation.  The following lemma is useful.

\blem\label{lemma-Nov23} (\cite{pavan})
Given two  matrices $W$ and $V$ of full column rank and of compatible dimension, define
\beq
\widehat{M} \triangleq  (W^\top W)^{-1} W^\top V (V^\top V)^{-1}V^\top W.
\eeq
Then
\beq
\mathrm{Im}(W)\cap \mathrm{Im}(V) = \{Wx: \widehat{M}x=x\}.
\eeq
\elem

 By Lemma \ref{lemma-Nov23}  we have the following result.
\bthm\la{thm_subspace_kernal}
We have
\bs\la{eq:gwgegewe_25nov}
\begin{align}
R_{c\bar{o}}=&\; {\rm Ker}\left(I+\mathbb{J}U_1U_1^\top\mathbb{J}U_2U_2^\top\right),
\la{eq:gwgegewe_25nov_cobar}
\\
 R_{co}=&\; {\rm Ker}\left(I+(\mathbb{J}U_1U_1^\top)^2\right),
 \la{eq:gwgegewe_25nov_co}
  \\
   R_{\bar{c}\bar{o}} = &\;    {\rm Ker}\left(I+(\mathbb{J}U_2U_2^\top)^2\right),
  \la{eq:gwgegewe_25nov_cbarobar}
  \\
   R_{\bar{c}o} = &\;   {\rm Ker}\left(I+\mathbb{J}U_2U_2^\top\mathbb{J}U_1U_1^\top\right).
   \la{eq:gwgegewe_25nov_cbaro}
\end{align}
\es
\ethm

\textbf{proof.} Notice that both $U_1$ and $U_2$ are of full column rank. Thus Lemma \ref{lemma-Nov23} is applicable. 
Firstly, we derive Eq. \eqref{eq:gwgegewe_25nov_co}.  Set $W=\mathbb{J}U_1$ and $V=U_1$. By Lemma \ref{lemma-Nov23}, 
\begin{equation}\label{nov25-4}
{\rm Im}(\mathbb{J}U_1)\cap{\rm Im}(U_1)=
\left\{Wx:x=U_1^\top\mathbb{J}^\top U_1U_1^\top \mathbb{J}U_1x\right\}.
\end{equation}
Inserting $W=\mathbb{J}U_1$ into the equality in Eq. \eqref{nov25-4}, yields that
\begin{equation}
Wx=\mathbb{J}U_1U_1^\top\mathbb{J}^\top U_1U_1^\top Wx
=-(\mathbb{J}U_1U_1^\top)^2 Wx.
\end{equation}
Thus, we have
\begin{equation}
\left(I+(\mathbb{J}U_1U_1^\top)^2\right) Wx =0,
\end{equation}
which  yields \eqref{eq:gwgegewe_25nov_co}. Secondly,  Eq.\eqref{eq:gwgegewe_25nov_cbarobar}  can be derived following the above procedure by replacing $U_1$ with $U_2$.
%
Thirdly, we derive Eq. \eqref{eq:gwgegewe_25nov_cobar}. Set $W=\mathbb{J}U_1$ and $V=U_2$. Then by Lemma \ref{lemma-Nov23},
\begin{equation}
{\rm Im}(\mathbb{J}U_1)\cap{\rm Im}(U_2)=
\left\{Wx:x=U_1^\top\mathbb{J}^\top U_2U_2^\top \mathbb{J}U_1x\right\}.
\end{equation}
Hence, we have 
\beq
Wx = -\mathbb{J}U_1U_1^\top\mathbb{J}U_2U_2^\top Wx,
\eeq
which leads to \eqref{eq:gwgegewe_25nov_cobar}. 
Finally, set $W=\mathbb{J}U_2$ and $V=U_1$, then by Lemma \ref{lemma-Nov23},  
\begin{equation}
{\rm Im}(\mathbb{J}U_2)\cap{\rm Im}(U_1)=
\left\{Wx:x=U_2^\top\mathbb{J}^\top U_1U_1^\top \mathbb{J}U_2x\right\}.
\end{equation}
Hence, we have 
\beq
Wx = -\mathbb{J}U_2U_2^\top\mathbb{J}U_1U_1^\top Wx,
\eeq
which leads to Eq. \eqref{eq:gwgegewe_25nov_cbaro}.
$\Box$

According to Theorem \ref{thm_subspace_kernal},  Hamiltonian matrices such as $\mathbb{J}U_1U_1^\top$ and skew-Hamiltonian matrices such as $(\mathbb{J}U_1U_1^\top)^2$ can be employed to characterize various subspaces. In what follows, we present some of  their properties.

\bcor\la{cor:lambda2}
The eigenvalues of the matrices $\mathbb{J}U_jU_j^\top \mathbb{J}U_kU_k^\top$ must be $0$ or $-1$ for all $j,k=1,2$. 
\ecor

\textbf{Proof.} 
We give the proof of the case of $\mathbb{J}U_1U_1^\top \mathbb{J}U_1U_1^\top$. The other cases can be proved similarly.
Notice that $\mathbb{J}U_1U_1^\top\mathbb{J}U_1U_1^\top U_2=0$,
which means that $0$ is the eigenvalue of the matrix $\mathbb{J}U_1U_1^\top\mathbb{J}U_1U_1^\top$ if ${\rm rank}(U_1) <2n$. On the other hand,  $\mathbb{J}U_1U_1^\top\mathbb{J}^\top U_1U_1^\top$ and $U_1^\top\mathbb{J}U_1U_1^\top\mathbb{J}^\top U_1$ share the same nonzero eigenvalues. Since the only nonzero eigenvalue of $U_1^\top\mathbb{J}U_1U_1^\top\mathbb{J}^\top U_1$ is $1$,  the nonzero eigenvalue of $\mathbb{J}U_1U_1^\top\mathbb{J}^\top U_1U_1^\top$ must be $1$. Moreover, it follows from $\mathbb{J}U_1U_1^\top\mathbb{J}U_1U_1^\top=-\mathbb{J}U_1U_1^\top\mathbb{J}^\top U_1U_1^\top$ that only $-1$ is the nonzero eigenvalue of $\mathbb{J}U_1U_1^\top\mathbb{J}U_1U_1^\top$.
$\Box$

The following result is an immediate consequence of Theorem \ref{thm_subspace_kernal} and   Corollary \ref{cor:lambda2}.

\bcor
We have
\bei
\item The controllable and unobservable subspace $R_{c\bar{o}}$ is spanned by the eigenvectors of  the matrix $\mathbb{J}U_1U_1^\top\mathbb{J}U_2U_2^\top$ associated with the eigenvalue $-1$.

\item The controllable and observable subspace $R_{co}$ is spanned by the eigenvectors of the skew-Hamiltonian matrix $(\mathbb{J}U_1U_1^\top)^2$ associated with the  eigenvalue $-1$.

\item The uncontrollable and unobservable subspace $R_{\bar{c}\bar{o}}$ is spanned by the  eigenvectors of the skew-Hamiltonian matrix $(\mathbb{J}U_2U_2^\top)^2$ associated with the eigenvalue $-1$. 

\item The uncontrollable and observable subspace $R_{\bar{c}o}$ is spanned by the eigenvectors of  the  matrix $\mathbb{J}U_2U_2^\top\mathbb{J}U_1U_1^\top$ associated with the eigenvalue $-1$.
\eei
\ecor

On one hand, as both $(\mathbb{J}U_1U_1^\top)^2$ and $(\mathbb{J}U_2U_2^\top)^2$  are skew-Hamiltonian matrices, there exist numerically stable algorithms for computing their eigenvectors; see for example \cite{vL84}, \cite{FMM+99}, \cite{Datta04}, \cite[Section 7.8]{GvL13}.  \footnote{Take the matrix $(\mathbb{J}U_1U_1^\top)^2$ as an example for concreteness. As $(\mathbb{J}U_1U_1^\top)^2$ is   skew-Hamiltonian, there exists a real orthogonal matrix $V$ such that 
\beq\la{eq:29nov_Schur}
V^\top (\mathbb{J}U_1U_1^\top)^2 V
=
\left[
\bey{cc}
M & N\\
0 & M^\top
\eey
\right],
\eeq
where $N$ is skew-symmetric and $M$ is a  quasi-triangular matrix whose diagonal blocks are either real scalars or $2 \times 2$ matrices. In general, these  $2 \times 2$ matrices have complex conjugate eigenvalues. Interestingly,  by Corollary \ref{cor:lambda2} we know that the eigenvalues of the matrix $(\mathbb{J}U_1U_1^\top)^2$ must be 0 or $-1$. As a result, $M$ must be an upper triangular matrix whose diagonal entries are $-1$ or 0. Thus, one may find a set of orthonormal eigenvectors of the matrix on the RHS of Eq. \eqref{eq:29nov_Schur} associated with the eigenvalue $-1$, and then left-multiply them by the real orthogonal matrix $V$ to get an orthonormal basis of the subspace $R_{co}$.
}

On the other hand, the matrix $\mathbb{J}U_1U_1^\top\mathbb{J}U_2U_2^\top$ is not skew-Hamiltonian. Certainly, we can get an orthonormal basis of $R_{c\bar{o}}$ by finding the eigenvectors of the matrix $\mathbb{J}U_1U_1^\top\mathbb{J}U_2U_2^\top$ associated with the eigenvalue $-1$. However, this matrix is a product of two matrices $\mathbb{J}U_1U_1^\top$ and $\mathbb{J}U_2U_2^\top$. For stable numerical computation, we may follow an alternative path as given below.



\bcor \la{cor:Linear_systems}
$x\in R_{c\bar{o}}$ if and only if 
\beq\la{eq:linear_sys_2}
x=U_2U_2^\top x = \bb[J]_n^\top U_1U_1^\top \bb[J]_n x.
\eeq
\ecor
\textbf{Proof.} 
Noticing
\beqm\la{eq:30nov_sharp}
\left(I+\mathbb{J}_n U_1U_1^\top\mathbb{J}_n U_2U_2^\top\right)^\sharp = I+\mathbb{J}_n U_2U_2^\top\mathbb{J}_n U_1U_1^\top,
\eeqm 
we have 
\beqnm
R_{\bar{c}o} 
&=&
{\rm Ker}( I+\mathbb{J}_n U_2U_2^\top\mathbb{J}_n U_1U_1^\top)
\nonumber\\
&=&
\mathbb{J}_n {\rm Ker}( I+ U_2U_2^\top\mathbb{J}_n U_1U_1^\top\mathbb{J}_n).
\eeqnm
However, by Eq. \eqref{eq:J_subspace} we have
$R_{c\bar{o}} =  \mathbb{J}_n R_{\bar{c}o} =\mathbb{J}_n^\top R_{\bar{c}o}$.
Hence,
\beqm
{\rm Ker}\left(I-\mathbb{J}_n^\top U_1U_1^\top \mathbb{J}_nU_2U_2^\top\right)
=
{\rm Ker}\left( I- U_2U_2^\top\mathbb{J}_n^\top U_1U_1^\top\mathbb{J}_n\right)
. 
\eeqm
This means that for all $x\in R_{c\bar{o}}$, 
\beqm
x = U_2U_2^\top\mathbb{J}_n^\top U_1U_1^\top\mathbb{J}_n x
=\mathbb{J}_n^\top U_1U_1^\top \mathbb{J}_nU_2U_2^\top x.
\eeqm
Consequently,
\beqnm
x&=& U_2U_2^\top\mathbb{J}_n^\top U_1U_1^\top\mathbb{J}_n x
\\
&=&
(U_2U_2^\top)(\mathbb{J}_n^\top U_1U_1^\top\mathbb{J}_n) (\mathbb{J}_n^\top U_1U_1^\top \mathbb{J}_n)(U_2U_2^\top) x
\\
&=&
(U_2U_2^\top) \mathbb{J}_n^\top U_1U_1^\top \mathbb{J}_nU_2U_2^\top x
\\
&=&
U_2U_2^\top x
\eeqnm
and
\beqnm
x &=&  \mathbb{J}_n^\top U_1U_1^\top \mathbb{J}_nU_2U_2^\top x
\\
&=&
(\mathbb{J}_n^\top U_1U_1^\top \mathbb{J}_n)(U_2U_2^\top)  (U_2U_2^\top)(\mathbb{J}_n^\top U_1U_1^\top\mathbb{J}_n) x
\\
&=&
(\mathbb{J}_n^\top U_1U_1^\top \mathbb{J}_n)  U_2U_2^\top\mathbb{J}_n^\top U_1U_1^\top\mathbb{J}_n x
\\
&=&
\mathbb{J}_n^\top U_1U_1^\top \mathbb{J}_n x.
\eeqnm
The proof is completed.
$\Box$


\bmrk
According to Corollary \ref{cor:Linear_systems},  to find an orthonormal basis of the subspace $R_{c\bar{o}}$, it suffices to solve the system of linear equations \eqref{eq:linear_sys_2} and then apply the Gram-Schmidt orthogonalization procedure.
\emrk



\section{Examples} \la{sec:example}


As an illustration, consider a class of $3$-mode, single-input-single-output (SISO) linear quantum systems  given in \cite{GZPG17}. The system Hamiltonian and coupling operator are 
\begin{equation}
\mbf{H}=\frac{\omega}{2}(\mbf{q}_3^2+\mbf{p}_3^2)+\lambda \mbf{q}_1\mbf{q}
_3+\lambda \mbf{q}_2\mbf{q}_3,
\end{equation}
and 
\begin{equation}
\mbf{L}=\frac{\gamma}{\sqrt{2}}(\mbf{q}_3+\imath \mbf{p}_3),
\end{equation}
respectively. Then, by Eq. \eqref{eq:real_sys_24nov} we have
\begin{equation}
\Lambda =\left[
\begin{array}{cccccc}
 0 & 0 & \frac{\gamma }{\sqrt{2}} & 0 & 0 & \imath \frac{\gamma }{\sqrt{2}} 
\end{array}
\right],
\end{equation}
and 
\begin{equation}
\mathbb{H}=\left[
\begin{array}{cccccc}
 0 & 0 & \lambda  & 0 & 0 & 0 \\
 0 & 0 & \lambda  & 0 & 0 & 0 \\
 \lambda  & \lambda  & \omega  & 0 & 0 & 0 \\
 0 & 0 & 0 & 0 & 0 & 0 \\
 0 & 0 & 0 & 0 & 0 & 0 \\
 0 & 0 & 0 & 0 & 0 & \omega  \\
\end{array}
\right].
\end{equation}



    


Based on the development in the previous sections, we find the following real orthogonal and block symplectic transformation matrix $\tilde{T}$ 
    \begin{equation}
\tilde{T}
=\left[\begin{array}{c|c|c}
 T_h & T_{co} & T_{\bar{c}\bar{o}}
\end{array}\right]
=
\left[
\begin{array}{cc|cc|cc}
 0 & \frac{1}{\sqrt{2}} & 0 & 0 & \frac{1}{2} & \frac{1}{2} \\
 0 & \frac{1}{\sqrt{2}} & 0 & 0 & -\frac{1}{2} & -\frac{1}{2} \\
 0 & 0 & -\frac{1}{\sqrt{2}} & -\frac{1}{\sqrt{2}} & 0 & 0 \\
 -\frac{1}{\sqrt{2}} & 0 & 0 & 0 & -\frac{1}{2} & \frac{1}{2} \\
 -\frac{1}{\sqrt{2}} & 0 & 0 & 0 & \frac{1}{2} & -\frac{1}{2} \\
 0 & 0 & \frac{1}{\sqrt{2}} & -\frac{1}{\sqrt{2}} & 0 & 0 \\
\end{array}
\right].
\end{equation}
Accordingly, the linear quantum  system takes the following Kalman canonical form \eqref{real_Kalman_ss}
with transformed coordinates
\begin{equation}\la{eq:x_bar}
\bar{\mbf{x}}=\left[
\begin{array}{c}
 -\frac{1}{\sqrt{2}}(p_1+p_2) \\ 
 \frac{1}{\sqrt{2}}(q_1+q_2) \\ \hline
 \frac{1}{\sqrt{2}}(p_3-q_3) \\
 -\frac{1}{\sqrt{2}}(p_3+q_3) \\ \hline
 \frac{1}{2}(-p_1+p_2+q_1-q_2) \\
 \frac{1}{2}(p_1-p_2+q_1-q_2) \\
\end{array}
\right], 
\end{equation}
corresponding to the ``$c\bar{o}$", ``$\bar{c}o$", ``$co$", and ``$\bar{c}\bar{o}$" system modes, respectively. It is easy to verify that system  matrices  are
\begin{equation}\begin{aligned}
&\bar{A}=\left[
\begin{array}{cc|cc|cc}
 0 & 0 & -\lambda  & -\lambda  & 0 & 0 \\
 0 & 0 & 0 & 0 & 0 & 0 \\
 \hline
 0 & -\lambda  & -\frac{\gamma ^2}{2} & \omega  & 0 & 0 \\
 0 & \lambda  & -\omega  & -\frac{\gamma ^2}{2} & 0 & 0 \\
 \hline
 0 & 0 & 0 & 0 & 0 & 0 \\
 0 & 0 & 0 & 0 & 0 & 0 \\
\end{array}
\right], \\
&\bar{B}=\left[
\begin{array}{cc}
 0 & 0 \\
 0 & 0 \\
  \hline
 \frac{\gamma }{\sqrt{2}} & -\frac{\gamma }{\sqrt{2}} \\
 \frac{\gamma }{\sqrt{2}} & \frac{\gamma }{\sqrt{2}} \\
  \hline
 0 & 0 \\
 0 & 0 \\
\end{array}
\right], ~~~~ \bar{C}=\left[
\begin{array}{cc|cc|cc}
 0 & 0 & -\frac{\gamma }{\sqrt{2}} & -\frac{\gamma }{\sqrt{2}} & 0 & 0 \\
 0 & 0 & \frac{\gamma }{\sqrt{2}} & -\frac{\gamma }{\sqrt{2}} & 0 & 0 \\
\end{array}
\right].
\end{aligned}\end{equation}
 Moreover, the Hamiltonian matrix under the coordinate $\bar{\mbf{x}}$ in this example can be calculated as
\begin{equation}
\bar{\mathbb{H}}=T^\top \mathbb{H} T=\left[
\begin{array}{cc|cc|cc}
 0 & 0 & 0 & 0 & 0 & 0 \\
 0 & 0 & -\lambda  & -\lambda  & 0 & 0 \\
 \hline
 0 & -\lambda  & \omega  & 0 & 0 & 0 \\
 0 & -\lambda  & 0 & \omega  & 0 & 0 \\
  \hline
 0 & 0 & 0 & 0 & 0 & 0 \\
 0 & 0 & 0 & 0 & 0 & 0 \\
\end{array}
\right],
\end{equation}
and by $\mbf{L}=\bar{\Lambda}\bar{\mbf{x}}$, where the complex matrix $\bar{\Lambda}$ satisfies
\begin{equation}\label{Gamma}
\left[\begin{array}{c}
\bar{\Lambda} \\
\bar{\Lambda}^{\#}
\end{array}
\right]
=
\left[
\begin{array}{cc|cc|cc}
0 & 0 & -\ff{1-\imath}{2}\gamma & -\ff{1+\imath}{2}\gamma & 0 & 0 \\
0 & 0 & -\ff{1+\imath}{2}\gamma & -\ff{1-\imath}{2}\gamma & 0 & 0
\end{array}
\right],
\end{equation}
which are consistent with the Kalman canonical forms derived in \cite[Eqs. (12)-(13)]{ZPL20}. Finally, the coordinate transformation 
\[
\breve{T}=\left[
\begin{array}{cccccc}
 0 & 0 & -\frac{1}{\sqrt{2}} & -\frac{1}{\sqrt{2}} & 0 & 0 \\
 0 & 1 & 0 & 0 & 0 & 0 \\
 0 & 0 & 0 & 0 & \frac{1}{\sqrt{2}} & \frac{1}{\sqrt{2}} \\
 0 & 0 & \frac{1}{\sqrt{2}} & -\frac{1}{\sqrt{2}} & 0 & 0 \\
 -1 & 0 & 0 & 0 & 0 & 0 \\
 0 & 0 & 0 & 0 & -\frac{1}{\sqrt{2}} & \frac{1}{\sqrt{2}} \\
\end{array}
\right] 
\]transforms $\bar{\mbf{x}}$ in Eq. \eqref{eq:x_bar} to 
\[
\breve{\mbf{x}}=\left[
\begin{array}{c}
 q_3 \\
 \frac{1}{\sqrt{2}}(q_1+q_2) \\
 \frac{1}{\sqrt{2}}(q_1-q_2) \\
 p_3 \\
 \frac{1}{\sqrt{2}}(p_1+p_2) \\
 \frac{1}{\sqrt{2}}(p_1-p_2) \\
\end{array}
\right],
\] 
and accordingly the system  is transformed to  the second Kalman canonical form as given by the last set of equations in  \cite{GZPG17}.


\section{Conclusion} \la{sec:con}
In this paper, we have presented a Gramian matrix approach to deriving the quantum Kalman canonical form for linear quantum systems in their real quadrature-operator representation. A detailed numerical computational procedure has also been provided for the construction of the real  orthogonal and block symplectic coordinate transformation matrices to transform a given quantum linear system to the Kalman canonical form.

\bibliographystyle{abbrv}

\end{document}